\documentclass[reprint,amssymb,twocolumn,noeprint,longbibliography]{revtex4-2}
\usepackage{graphicx}
\usepackage{amsfonts}
\usepackage{amsmath}
\usepackage[usenames,dvipsnames]{color}
\usepackage{bm}
\usepackage{amssymb}

\usepackage{multirow}
\usepackage{epstopdf}
\usepackage{natbib}
\usepackage[normalem]{ulem}

\usepackage{hyperref}
\hypersetup{
colorlinks=true,
linkcolor=blue,
citecolor=blue,
filecolor=green,
urlcolor=blue,
}

\usepackage[monochrome]{xcolor}

\def\k1{k_1}
\def\k2{k_2}
\def\q1{q_1}
\def\q2{q_2}

\def\({\left (}
\def\){\right )}
\def\[{\left [}
\def\]{\right ]}

\newcommand{\beq}{\begin{equation}}
\newcommand{\eeq}{\end{equation}}

\begin{document}
\date{\today}
\flushbottom \draft
\title{Universal stereodynamics  of cold atom-molecule collisions in electric fields} 

\author{Timur V. Tscherbul$^{1}$ and Jacek K{\l}os$^{2}$}

\affiliation{$^{1}$Department of Physics, University of Nevada, Reno, NV, 89557, USA\\
$^{2}$Department of Physics, Joint Quantum Institute, University of Maryland, College Park, Maryland, 20742, USA}

\begin{abstract}
 We use numerically exact quantum dynamics calculations to demonstrate universal  stereoselectivity  of cold  collisions of $^2\Pi$ molecules with $^1$S-state atoms in an external electric field. We show that cold collisions of OH molecules in their low-field-seeking $f$-states, whose dipole moments are oriented against the field direction, are much more likely to lead to inelastic scattering than those of molecules oriented along the field direction,
causing nearly perfect steric asymmetry in the inelastic collision cross sections.
      The universal nature of this effect is due to the threshold suppression of  inelastic scattering between the degenerate $\pm M$ Stark sublevels of the high-field-seeking $e$-state, where $M$ is the projection of the total angular momentum of the molecule  on the field axis.  
Above the $\Lambda$-doublet threshold, the stereodynamics of inelastic atom-molecule collisions  can be  tuned  via electric-field-induced resonances, which enable effective  control of  Ne~+~OH scattering  over the range of collision energies  achievable in current merged beam experiments. 
\end{abstract}



\maketitle
\clearpage
\newpage

\section{Introduction}

Modern experimental studies of ultracold molecular gases \cite{Bohn:17} have reached an 
  extraordinary level of control over molecular degrees of freedom using external electromagnetic fields \cite{Carr:09,Lemeshko:13,Krems:18}.   
  In particular, control over the rotational motion \cite{Koch:19} makes it possible to address a central question of chemical physics concerning the role of the relative orientation of the reactants in determining the outcome of molecular collisions and chemical reactions \cite{Herschbach:06,Leuken:96,Schreel:97,Beek:00,Lange:99,Alexander:00,Lange:04,Aquilanti:05,Liu:16,Wang:11,Wang:12,Liu:16,Klos:21}.   The steric effects have been the subject of numerous experimental studies in crossed molecular beams \cite{Leuken:96,Lange:99,Beek:00,Alexander:00,Lange:04,Cireasa:05,Aquilanti:05,Liu:16}, external field traps  \cite{Ni:10,Miranda:11}, and, more recently, in merged molecular beams \cite{Perreault:17,Perreault:18}. The latter experiments probed  the stereodynamics of cold HD~+~D$_2$ collisions at  1~K \cite{Perreault:17,Perreault:18}  and observed a dramatic preference for the perpendicular alignment of collision products.  The single partial-wave regime accessed in these experiments is optimal for studying and controlling collision stereodynamics \cite{Aldegunde:06} due to the absence of detrimental averaging over many partial waves  (or impact parameters), which tends to obscure steric effects \cite{Herschbach:06}.
  

  
  
  
Recent quantum scattering calculations revealed the important role of single scattering resonances in determining the stereodynamics of  cold HD$(v=1,j=2)$~+~H$_2$ collisions \cite{Croft:18} and suggested the possibility of tuning shape resonances in cold HD$(v=1,j=2)$~+~H$_2$ collisions by aligning the rotational angular momentum of HD with respect to the initial relative velocity vector \cite{Jambrina:19}. 
Additional calculations explored the stereodynamics of cold rotationally inelastic He~+~HD \cite{Morita:20a} and HCl~+~H$_2$ collisions \cite{Morita:20b}  in the presence of overlapping resonances and identified a universal trend in the stereodynamic preference of state-to-state integral cross sections.


Previous theoretical work on steric effects in cold molecular collisions  \cite{Croft:18,Jambrina:19,Morita:20a,Morita:20b} has focused on molecules in nondegenerate electronic states of $\Sigma$ symmetry in the absence of external fields. Open-shell molecular radicals such as OH$(^2\Pi)$ and NO$(^2\Pi)$ are readily controllable by external fields due to their quasi-degenerate $\Lambda$-doublet levels of the opposite parity \cite{Brown:03}. The OH radical was among the first molecules cooled and trapped at low temperatures \cite{Meerakker:06,Stuhl:14} and its cold collisional properties with rare-gas atoms have been extensively studied \cite{Meerakker:06,Stuhl:14,Kirste:10,Scharfenberg:11,Sawyer:11,Kirste:12}. External fields orient or align the molecules along a laboratory-fixed quantization axis  \cite{Stapelfeldt:03,Lemeshko:13,Friedrich:92,Gonzalez-Ferez:04,Lemeshko:08},  providing an extra spatial direction for observing novel stereodynamical effects. In addition, external fields are commonly used to tune the scattering properties of cold atoms and molecules via Feshbach resonances \cite{Chin:10}. A combination of steric and external field control may thus lead to  new and powerful ways to engineer the quantum dynamics of molecular collisions at ultralow temperatures.

Here, we explore the stereodynamics of cold atom-molecule collisions in an electric field using Ne~+~OH as a representative example (rare gas - OH collisions serve as prototype systems for studying steric effects in molecular collisions \cite{Leuken:96,Lange:99,Beek:00,Alexander:00,Lange:04}).  Using rigorous quantum scattering calculations, we uncover a universal  stereodynamical trend: Collisions of $^2\Pi$ molecules  initially oriented against the field direction are much more likely to lead to inelastic scattering than those of molecules oriented along the field direction. 
\textcolor{blue}{This is the first theoretical study of cold  atom-molecule collision stereodynamics in the presence of external fields using numerically exact quantum dynamics calculations.}
We also show that the stereodynamics of cold atom-molecule collisions can be controlled by an external electric field, and find  that such control can be extensive even in the multiple partial wave regime,
 which can be reached experimentally in merged molecular beams  \cite{Lavert-Ofir:14,Gordon:18}.
  Our predictions can thus be verified in current molecular beam scattering experiments.

\section{Theory} 

  To explore the stereodynamics of cold atom-molecule collisions in an external electric field, we carry out rigorous quantum  scattering calculations for the benchmark collision system Ne~+~OH parametrized by accurate {\it ab initio} interaction potentials as described in the Appendix.  
The stereodynamical observables of interest are encoded in
the atom-molecule scattering amplitude \cite{Krems:04,Tscherbul:08a}
\begin{equation}\label{ScatteringAmplitude}
q_{\alpha\to\alpha'} (\hat{k}_\alpha,\hat{R}) = 2\pi \sum_{\substack{ \ell, m_\ell, \\ \ell',m_\ell'} }
{i}^{\ell-\ell'} Y^{*}_{\ell m_\ell}(\hat{k}_\alpha)Y_{\ell' m_\ell'}(\hat{R})T_{\alpha \ell m_\ell;\alpha' \ell' m_\ell'}
\end{equation} 
where $\mathbf{k}_\alpha=k_\alpha\hat{k}_\alpha$ the incident wavevector,  $\hat{k}_\alpha$ gives the  direction of the incident flux with respect to the space-fixed (SF) $Z$-axis defined by the direction of the external electric field, $\hat{R}$ specifies the direction of the scattered flux,   $Y_{lm}(\hat{R})$ are the spherical harmonics, $\ell$ and $m_\ell$ are the quantum numbers for the orbital angular momentum and its SF projection,  $\alpha$ refers to the internal states of the molecule, and $T_{\gamma \ell m_\ell;\gamma' \ell' m_\ell'}$ are the transition $T$-matrix elements.

We will assume that the external field is collinear with the incident relative velocity vector ($Z\parallel \hat{k}_{\alpha}$) \cite{Croft:18,Jambrina:19,Morita:20a,Morita:20b} 
 which allows us to set $\hat{k}_\alpha=0$ in Eq. (\ref{ScatteringAmplitude}) to yield \cite{Tscherbul:08a}
\begin{equation}\label{ScatteringAmplitude0}
q_{\alpha\to\alpha'}^{(0)} (\hat{R}) = \sqrt{\pi} \sum_{ \ell, \ell',m_\ell'}
{i}^{\ell-\ell'} (2\ell+1)^{1/2} Y_{\ell' m_\ell'}(\hat{R})T_{\alpha \ell 0;\alpha' \ell' m_\ell'}
\end{equation} 
The integral cross section (ICS) corresponding to a fixed orientation of the incident flux ($\hat{k}_\alpha=\hat{0}$) may be obtained by integrating the  differential cross section ${d\sigma_{\alpha\to\alpha'} (\hat{R})}/{d\Omega} = {k_\alpha^{-2}}|q^{(0)}_{\alpha\to\alpha'} (\hat{R})|^2$ over all angles. Substituting the scattering amplitude from Eq. (\ref{ScatteringAmplitude0}) and performing the integration, we obtain \cite{Tscherbul:08a}
\begin{multline}\label{ICSsteric}
\sigma_{\alpha \to \alpha'}^{(0)}=\frac{\pi}{k_{\alpha}^2}\biggl{[}
\sum_{\ell}\sum_{\ell' m_\ell'}(2\ell + 1)|T_{\alpha \ell 0;\alpha' \ell' m_\ell'}|^2 \\ +
\sum_{\substack{  \ell_1\ne \ell_2 \\ \ell',m_\ell'} }[(2\ell_1+1)(2\ell_2+1)]^{1/2} {i}^{\ell_2 - \ell_1}
T^*_{\alpha \ell_1 0;\alpha' \ell' m_\ell'} T_{\alpha \ell_2 0;\alpha' \ell' m_\ell'} \biggr{]}.
\end{multline}
The first term on the right represents the incoherent contribution to the ICS, which does not depend on the phases of  $T$-matrix elements. The second term is an interference term, which originates from
 fixing the direction of the incident collision flux in Eq. (\ref{ScatteringAmplitude0}), as required for the description of  molecular beam stereodynamics experiments \cite{Croft:18,Jambrina:19,Morita:20a,Morita:20b}. 
  The ``steric'' ICS defined by Eq. (\ref{ICSsteric}) is notably different from the conventional state-to-state ICS $\sigma_{\alpha\to\alpha'}=\pi k_{\alpha}^{-2}\sum_{\ell,m_\ell}\sum_{\ell',m_\ell'} |T_{\alpha \ell m_\ell;\alpha' \ell' m_\ell'}|^2$ obtained by averaging the absolute square of the  scattering amplitude (\ref{ScatteringAmplitude}) over $\hat{k}_\alpha$ and integrating over $\hat{R}$ \cite{Krems:04} leading to the  disappearance of interference terms. 
  We note that the steric ICS (\ref{ICSsteric}) becomes identical to the conventional ICS in the $s$-wave limit, where $\ell_1=\ell_2=0$ \cite{Tscherbul:08a} and both types of ICS obey the same threshold laws for $s$-wave scattering \cite{Krems:03}. 
We will omit the prefix ``steric'' when referring to the ICS (\ref{ICSsteric}) unless necessary to avoid confusion.

The $T$-matrix elements in Eqs. (\ref{ScatteringAmplitude0})--(\ref{ICSsteric}) are obtained using numerically exact quantum scattering methodology  \cite{Tscherbul:09,Pavlovic:09} based on the accurate {\it ab initio} Ne-OH PESs  calculated as described in the Appendix. Unlike previous theoretical studies \cite{Leuken:96,Lange:99,Beek:00,Alexander:00,Lange:04}, our calculations explicitly account for the effects of external electric fields on quantum dynamics \cite{Tscherbul:09,Pavlovic:09} as required for the proper theoretical description of cold atom-molecule collisions \cite{Balakrishnan:16,Tscherbul:18c}.

\section{Results and discussion} 
 
\begin{figure}[t]
\includegraphics[width=0.94\linewidth, trim = 30 20  0 40]{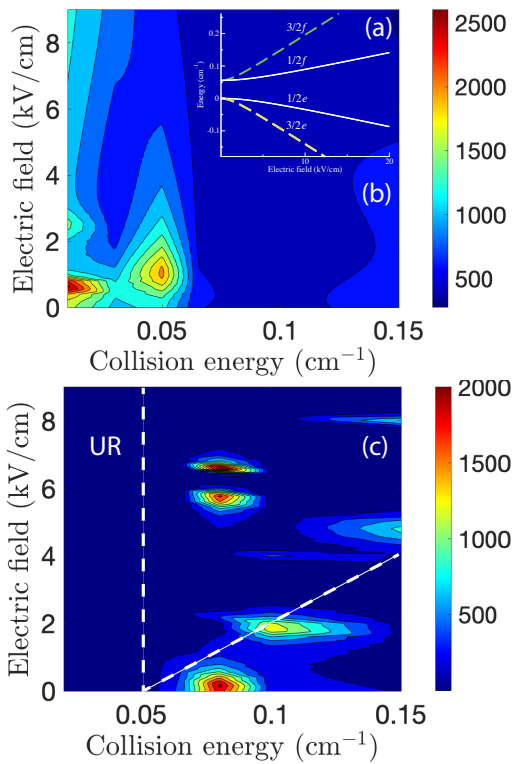}
\caption{(a) Stark energy levels of OH($^2\Pi$) in the $J=3/2$ ground-state manifold as a function of electric field \textcolor{blue}{[panel (a) is the inset in panel (b)]}. The initial states used in steric asymmetry computations are shown by dashed lines. Panels (b) and (c) show the ICSs (in units of $a_0^2$) for Ne~+~OH   plotted vs. collision energy and electric field  for the $|f,M=3/2\rangle$ (b) and $|e,M=3/2\rangle$ (c) initial states of OH. The zero-field $\Lambda$-doublet energy is marked by the vertical dashed line and the dependence $\Delta E_\Lambda(E)$ is marked by the slanted line. The area labeled ``UR'' indicates the universal threshold regime, where the inelastic ICS $\sigma^\text{inel}_{f,M}$ is suppressed. }
\label{fig:xsinel}
\end{figure}

Figure \ref{fig:xsinel}(a) shows the Stark energy levels of OH($X^2\Pi$)  in its ground vibronic state.
 The ground-state $J=3/2$ level is split by the spin-orbit interaction between the ground and excited electronic states of OH into a $\Lambda$-doublet consisting of two closely lying levels of opposite  parity \cite{Brown:03, Maeda:15}. The low-field seeking $M$-components of the upper $f$-state ($|M|=1/2$ and $|M|=3/2$) increase in energy with increasing electric field. The OH molecules residing in the $f$ states are oriented against the direction of the applied electric field \cite{Leuken:96}.  In contrast, the energy of the high-field-seeking Stark sublevels of the lower $e$-state decreases with increasing field as their dipole moments are oriented along the field direction \cite{Leuken:96}. 
 
In  Fig. \ref{fig:xsinel}(b) we show the \textcolor{blue}{total} inelastic  ICS $\sigma^\text{inel}_{i}=\sum_{k\ne i} \sigma_{i\to k}^{(0)}$ \textcolor{blue}{(hereafter referred to simply as the ICS, with the ICS  type explicitly indicated only when necessary to avoid confusion)} for the highest-energy low-field seeking Stark state $|i\rangle=|f,M=3/2\rangle$ of OH($J=3/2$) as a function of collision energy and electric field.
 We observe two pronounced  resonance peaks in the collision energy ($E_\text{coll}$) dependence of the ICS near $E=1$~kV/cm. These resonances are due to the trapping of the collision partners behind  centrifugal barriers in either the incoming or outgoing collision channels
   \cite{Pavlovic:09}. 
At $E_\text{coll}> 0.1$~cm$^{-1}$, Ne~+~OH collisions occur  in the multiple partial-wave regime, where destructive interference between different partial wave contributions washes out the resonance structure in the ICS apparent at lower collision energies.


 Figure \ref{fig:xsinel}(c) shows the ICS for the ground high-field-seeking state of OH  $|e,M=3/2\rangle$ colliding with Ne.  At low collision energies ($E_\text{coll} < \Delta E_\Lambda$, where $\Delta E_\Lambda=0.055$~cm$^{-1}$ is the $\Lambda$-doublet splitting energy)  inelastic scattering of OH molecules oriented along the electric field direction  is strongly suppressed, leading to the universal steric preference phenomenon considered below.
The origin of the suppression is that only three inelastic channels ($|e,M'=-3/2\rangle$ and $|e,M'=\pm1/2\rangle$) remain open at zero collision energy in the low field limit. As these channels have $M'\ne M$  inelastic scattering must be accompanied by a change in $M$. According to the  Krems-Dalgarno threshold laws  \cite{Krems:03} transitions with nonzero $\Delta M$ scale with the collision energy as $E_\text{coll}^{\Delta M}$ (for even $\Delta M$) and $E_\text{coll}^{\Delta M+1}$ (for odd $\Delta M$). Thus, transitions with $\Delta M\ne 0$ will be suppressed in the $s$-wave limit by centrifugal barriers in the outgoing collision channel.

 At higher collision energies above the $\Lambda$-doublet threshold, inelastic channels of opposite parity and $\Delta M=0$ (such as $|f,M'=3/2\rangle$) open up, causing a substantial increase of the ICS and the appearance of near-threshold resonances.
  A total of 8 isolated resonances  occur over the range of collision energies 0.06-0.2 cm$^{-1}$ below 9 kV/cm. Remarkably, the resonances survive not only in the few partial wave regime ($E_\text{coll}<0.05$~cm$^{-1}$) but also at higher collision energies. This is due to the limited number of inelastic channels available for the $|e,M=3/2\rangle$ initial state to decay into, resulting in a more pronounced $S$-matrix pole structure compared to the $|f,M=3/2\rangle$ initial state \cite{Hutson:07}.
  
  
The resonance structure shown in Fig.~\ref{fig:xsinel}(c)  displays an interesting pattern, shifting to higher collision energies with increasing field. This occurs due to the field-induced repulsion between the opposite parity $e$ and $f$  states.  As a result of widening   $\Lambda$-doublet  energy gap, the minimum collision energy required to access the inelastic channels in the $f$-manifold increases with the $E$-field, shifting the onset of the resonance pattern to higher collision energies.

Additionally, as seen in Fig.~\ref{fig:xsinel}(c), increasing the electric field {\it suppresses} inelastic scattering from the $|e,M=3/2\rangle$ state of OH:  the resonance  maxima of the ICS become less pronounced at higher electric fields. 
This is caused by the energy gap between the $|M|=1/2$ and $|M|=3/2$ states in the $e$-manifold growing linearly with increasing field [see Fig.~\ref{fig:xsinel}(a)] until the $|M|=1/2$ channels become closed. The ICS for the single energetically allowed   transition  $|e,M=3/2\rangle \to |e,M'=-3/2\rangle$ scales as $\sigma^\text{inel}\simeq E_\text{coll}^{\Delta M +1} = E_\text{coll}^4$ \cite{Krems:03} and is thus strongly suppressed at ultralow collision energies. As shown below, the suppression of the ICS is a {\it universal trend}, which manifests itself in ultracold collisions of $^2\Pi$ molecules in the $|e,M=3/2\rangle$ initial state  with spherically symmetric atoms. In contrast, no such trend exists for the $|f,M=3/2\rangle$ initial state.

\begin{figure}[t]
\includegraphics[width=1.1\linewidth, trim = 70 160  0 85]{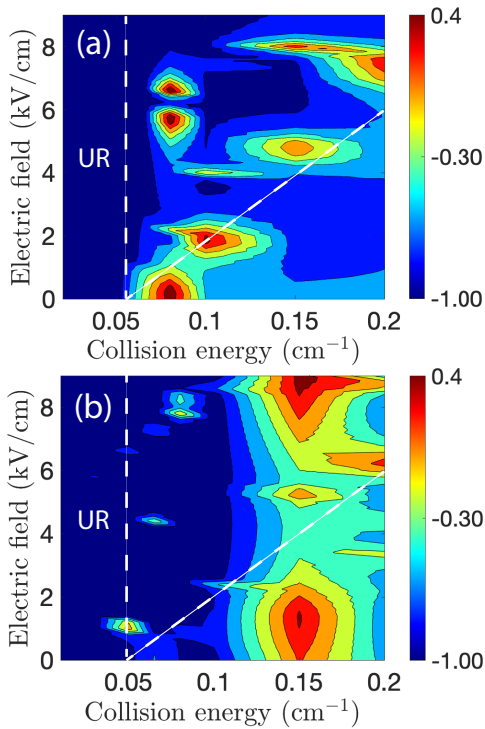}
\caption{(a) Steric asymmetry (\ref{steric_asymm}) for inelastic Ne~+~OH collisions as a function of  collision energy and electric field computed using the present Ne-OH PESs (a) and the PESs from Ref.~\cite{Sumiyoshi:10} (b) [see the Appendix for details].  The $\Lambda$-doublet splitting energy of OH is marked by the vertical dashed line ($E=0$) and by the sloped dashed line (as a function of $E$). }
\label{fig:steric_asym}
\end{figure}

To gain additional insight into cold Ne~+~OH collision stereodynamics,  we calculate the steric asymmetry parameter \cite{Leuken:96,Lange:99,Beek:00,Alexander:00,Lange:04}
\begin{equation}\label{steric_asymm}
\mathcal{S}^\text{inel} =  \frac{\sigma_{e,M}^\text{inel} - \sigma_{f,M}^\text{inel}}{\sigma_{e,M}^\text{inel} + \sigma_{f,M}^\text{inel} }
\end{equation}
where $\sigma_{i}^\text{inel} $  is the ICS for the \textcolor{blue}{$i$-th} initial state of OH aligned along  and against the field axis (see above), $M=3/2$, and  the $k$ sum runs over all energetically accessible final channels.  The steric asymmetry measures the difference between the collisional properties of OH molecules oriented along vs. against the field direction (with coincides with the incident atom-molecule velocity vector). A value of $\mathcal{S}^\text{inel}$ close to -1 (+1) indicates a strong stereodynamic preference for Ne~+~OH  collisions with OH oriented against (along) the field axis. 
Experimental measurements and theoretical calculations of the steric asymmetry have provided a wealth of valuable information about stereodynamic effects  in collisions of OH and NO molecules with rare-gas atoms at collision energies of 300~K and above \cite{Leuken:96,Lange:99,Beek:00,Alexander:00,Lange:04}.
\color{blue}
 In particular, Schreel and Meulen \cite{Schreel:97} and Beek {\it et al.} \cite{Beek:00} measured and calculated the steric asymmetry of rotationally inelastic collisions of   OH($J=3/2,f$) molecules with  He, Ar, and H$_2$ at  collision energies $E_\text{coll}=394{-}746$ cm$^{-1}$. They observed steric asymmetries ranging from $\mathcal{S}^\text{inel} =-0.01$ to $-0.27$ depending on the collision energy and the rotational transition (the largest asymmetry was observed for the $J=3/2,f \to 7/2, e$ transition in He~+~OH collisions at $E_\text{coll} =394$ cm$^{-1}$ \cite{Schreel:97}.) Larger steric asymmetries (up to $|\mathcal{S}^\text{inel}| = 0.78$) were reported for rotationally inelastic Ar~+~OH collisions at $E_\text{coll}=746$~cm$^{-1}$ \cite{Beek:00}.   Interestingly, the preference for the negative steric asymmetry observed in high-temperature He~+~OH collisions  \cite{Schreel:97}  is also apparent in our Ne~+~OH calculations  (see Fig.~\ref{fig:steric_asym}). 
 
 It should be noted, however, that previous experimental and theoretical work on He~+~OH and Ar~+~OH collisions \cite{Schreel:97,Beek:00} was performed at much higher collision energies ($E_\text{coll}=394{-}746$ cm$^{-1}$) and for a different type of inelastic process (rotationally inelastic collisions) than explored here. Under the  conditions of Refs.~\cite{Schreel:97,Beek:00} the stereodynamical properties of atom-molecule collisions are determined by contributions from many partial waves.  Because of these multiple competing contributions, the physical mechanisms responsible for, e.g., the negative sign of the steric asymmetry in high-temperature He~+~OH collisions \cite{Schreel:97,Beek:00} are significantly more challenging to identify and interpret than the low-temperature trends explored here.



 \color{black}

Figure \ref{fig:steric_asym}(a) is  a two-dimensional map of the steric asymmetry for inelastic Ne~+~OH collisions plotted as a function of collision energy and electric field. Two distinct regions  may be observed in the map, which we will refer to as the universal region and the resonance region. The universal region corresponds to the regime $E_\text{coll}<\Delta E_\Lambda$, where  the upper components of the $\Lambda$-doublet are closed, and inelastic scattering from the lowest low-field-seeking state $|e,J=3/2,M=3/2\rangle$ of OH is strongly suppressed by the threshold laws for $M$-changing collisions as noted above.
Thus, in the universal region, $\sigma_{f,M}^\text{inel} \gg \sigma_{e,M}^\text{inel}$ and $\mathcal{S}^\text{inel}\simeq -1$.  This remarkable trend in only apparent in the {inelastic} ICSs as further discussed in the Appendix.

Significantly, the universal suppression of inelastic scattering persists at nonzero electric fields because the  $M=3/2$ and $M=-3/2$ magnetic sublevels remain degenerate, and thus zero-field Krems-Dalgarno threshold laws \cite{Krems:03} remain applicable to the $|e,M=3/2\rangle\to |e,M=-3/2\rangle$ transition. The degeneracy can be lifted by an external magnetic field, which is expected to break the universal behavior of the steric asymmetry, leading to a rapid increase of $\mathcal{S}^\text{inel}$ as the $s$-wave threshold scaling of the $M$-changing ICS changes from $~E_\text{coll}^4$ to $E_\text{coll}^{-1/2}$. 

In the resonance regime defined by the condition $E_\text{coll}>\Delta E_\Lambda$ excitation transitions occur from the initial $e$-state, such as  $|e,M=3/2 \to |e,M=\pm1/2 \rangle$ and $|e,M=3/2 \to |f,M=\pm1/2 \rangle$. While these are also $M$-changing transitions, they are not so strongly suppressed compared to the $|e,M=3/2\rangle\to |e,M=-3/2\rangle$ transition due to their smaller $\Delta M$. As s result, the background value of the ICS increases and so does the steric asymmetry. In addition, scattering resonances begin to appear near and above the excitation thresholds leading to distinct spikes in  $\sigma_{e,M}^\text{inel}$ [see Fig.~\ref{fig:xsinel}(c)]. As noted above and seen in Fig.~\ref{fig:xsinel}(b),  $\sigma_{f,M}^\text{inel}$ does not vary strongly with either collision energy or electric field in the resonant regime. Taken together, these factors cause the appearance of the resonance peaks in the steric asymmetry in Fig.~\ref{fig:steric_asym}(a).  As the details of the resonance structure are sensitive to the underlying PESs (a well-documented phenomenon in cold molecular collisions \cite{Hutson:07,Suleimanov:16,Morita:19}), the resonance regime can also  be regarded as nonuniversal.  \textcolor{blue}{The significance of the nonuniversal behavior is that it allows one to control the stereodynamics of cold atom-molecule collisions by applying an external electric field. For example, as shown in Fig.~\ref{fig:steric_asym}, by varying the external electric field between zero and 10~kV/cm (an interval well within the reach of modern experimental capabilities), it is possible to tune the steric asymmetry over a substantial range from $-1$ to 0.4, essentially reversing the stereodynamical preference of inelastic  Ne + OH collisions.  The nonuniversal behavior could be observed as a function of collision energy and external fields in merged beam experiments with state-selected OH molecules and Ne atoms  as was done for cold He$^*$~+~H$_2$ and Ne$^*$~+~Ar collisions \cite{Lavert-Ofir:14,Gordon:18}. }

\begin{figure}[t]
\includegraphics[width=1.05\linewidth, trim = 50 170  0 162]{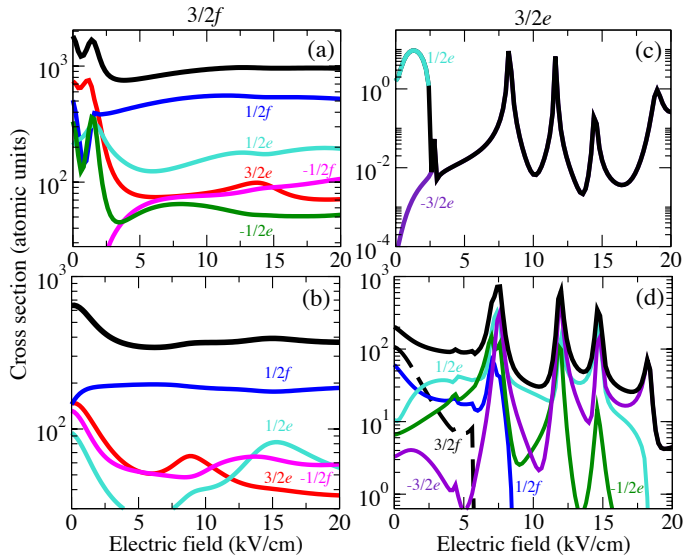}
\caption{(a) Electric field dependence of state-to-state ICSs $\sigma_{f,M=3/2\to k}$ [panels (a), (b)] and $\sigma_{e,M=3/2\to k}$ [panels (c), (d)] for Ne~+~OH collisions.  The collision energy is 0.02 cm$^{-1}$ [panels (a), (c)] and 0.2 cm$^{-1}$ [panels (b),  (d)]. The ICSs summed over all final $k$ are shown by the top traces.}
\label{fig:state-to-state}
\end{figure} 

To illustrate the distinction between the universal vs. nonuniversal  regimes, we plot in Fig.~\ref{fig:steric_asym}(b) the steric asymmetry calculated using a different set of Ne-OH interaction PESs   \cite{Sumiyoshi:10}. While the differences between the PESs are small (see the Appendix), they have a dramatic effect  on the resonance structure  at $E_\text{coll}\ge \Delta E_\Lambda$, as expected in the nonuniversal regime \cite{Hutson:07,Suleimanov:16,Morita:19}.

Remarkably, by comparing Fig.~\ref{fig:steric_asym}(a) and Fig.~\ref{fig:steric_asym}(b) we observe  that the steric asymmetries calculated using the different PESs are in nearly perfect agreement with each other at  $E_\text{coll} < \Delta E_\Lambda$  ($\mathcal{S}^\text{inel}\simeq -1$). This independence is a clear signature of the universal regime, where inelastic scattering of OH($|e,M=3/2\rangle$) is strongly suppressed due to the threshold effects \cite{Krems:03} (see above).

In Figs.~\ref{fig:steric_asym}(a) and \ref{fig:steric_asym}(b) we observe small deviations from the universal behavior at $E_\text{coll} \simeq \Delta E_\Lambda$. 
To understand the origin of these deviations, we plot in Fig.~\ref{fig:state-to-state} the state-to-state ICSs  $\sigma_{f,M=3/2\to k}$ and  $\sigma_{e,M=3/2\to k}$, which define the steric asymmetry (\ref{steric_asymm}). Below the $\Lambda$-doublet threshold the electric field dependence of $\sigma_{f,M=3/2\to k}$ is determined by isolated shape resonances  with the dominant contribution due to the  $|e,M=3/2\rangle$ final state.  At higher electric fields and/or collision energies, the resonances  broaden  and begin to overlap, leading to the disappearance of distinct peaks in the total ICS \cite{Pavlovic:09}.  We note that the resonance structure in the state-to-state ICS can survive at collision energies as high as 0.2 cm$^{-1}$, as illustrated in Fig.~\ref{fig:state-to-state}(b) for the final state $|e,M=1/2\rangle$.


As shown in Figs.~\ref{fig:state-to-state}(c) and \ref{fig:state-to-state}(d),  the ICSs $\sigma_{e,M=3/2\to k}$ increase by 2-4 orders of magnitude with increasing collision energy by a factor of 10, which is consistent with their $ E_\text{coll}^{\Delta M}$  threshold scaling discussed above \cite{Krems:03}. In contrast, the ICSs $\sigma_{f,M=3/2\to k}$ {\it decrease} due to their different threshold scaling $\simeq E_\text{coll}^{-1/2}$. We  verified that the deviations from the perfect  universal scaling ($\mathcal{S}^\text{inel}=-1$) occur due to the small contributions of the $\ell \ge 1$ partial waves to the  ICS $\sigma_{e,M=3/2\to k}$ at $E_\text{coll} < \Delta E_\Lambda$ [see Fig.~\ref{fig:state-to-state}(c)]. At lower collision energies, these contributions freeze out as $E_\text{coll}^4$ and the universal relation $\mathcal{S}^\text{inel}=-1$ holds to an increasingly better accuracy. 


\section{Summary and outlook} 

In summary, we have established a universal stereodynamical trend in  cold collisions of $^2\Pi$ molecular radicals with $^1\text{S}_0$-state atoms in an external electric field. Using rigorous quantum scattering calculations based on highly accurate {\it ab initio} interaction potentials, we show that the steric anisotropy of Ne~+~OH collisions approaches $-1$ in the limit of zero collision energy due to a suppression of  $M$-changing transitions from the $|e,M=3/2\rangle$ initial state, in which the dipole moment of OH is oriented along the field direction. The suppression occurs universally in the  $s$-wave threshold regime, where the $M$-changing cross sections vanish \cite{Krems:03}, and it persists at collision energies below the $\Lambda$-doublet energy $\Delta E_\Lambda$ regardless of the magnitude of the applied electric field [see Fig.~2]. Above the $\Lambda$-doublet energy, nearly perfect stereoselectivity is lost and scattering occurs in the nonuniversal resonant regime, where extensive control is possible over collision stereodynamics via electric field-induced resonances. Remarkably, the resonances survive  at collision energies as high as 0.2 cm$^{-1}$ [see Fig.~2], which can be realized in merged molecular beams \cite{Lavert-Ofir:14,Gordon:18}, making our predictions verifiable in current cold molecule experiments.

\section*{Acknowledgements} 
We are grateful to Jun Ye, Gerrit Groenenboom, Hao Wu, David Reens, and Piotr Wcis{\l}o  for stimulating discussions, and to Yoshihiro Sumiyoshi for sharing the Ne-OH PESs \cite{Sumiyoshi:10}.
This work was supported by the NSF EPSCoR RII Track-4 Fellowship (Award No. 1929190).

\section*{Appendix A: Ab initio calculations of  Potential Energy Surfaces}



In this Appendix we provide the details of our {\it ab initio} calculations of Ne-OH($^2\Pi$) interaction potentials. Our  quantum scattering calculations based on these potentials are described in Sec. B below.

We performed electronic structure calculations of the adiabatic potential energy surfaces (PESs) for the Ne-OH(X$^2\Pi$) van der Waals complex at the explicitly correlated coupled cluster level of theory. Due to the fact that the OH molecule is a spin doublet radical,  we used the spin-restricted version of the coupled cluster method with single, double and perturbative triple corrections (RCCSD(T)-F12a) using the F12a ansatz with scaled triples corrections as implemented in the MOLPRO program~\cite{MOLPRO}. The singly occupied molecular orbital of  OH(X$^2\Pi$) splits its degeneracy upon approach of the Ne atom to form the electronic states of the Ne-OH complex. The electronic wavefunctions of the complex transform according to either $A'$  or $A''$ irreducible representations of the C$_s$ symmetry group, which are either symmetric or antisymmetric with respect to reflection in the triatomic plane (for non-collinear geometries) and according to the $\Pi$ representation of the C$_{\infty v}$ point group in collinear geometries. 

The initial single reference wavefunctions are obtained from spin-restricted Hartree-Fock (RHF) calculations, which control the corresponding $A'$ or $A''$ symmetries of the wavefunctions of the Ne-OH dimer and OH monomer  for subsequent RCCSD(T)-F12 calculations.  We use augmented correlation-consistent triple-zeta atomic basis sets (aug-cc-pVTZ)~\cite{dunning:89} with automatically generated auxiliary density fitting sets for the explicitly-correlated part of the calculations. We additionally augment the basis with mid-bond functions that improve the description of the dispersion-bound complexes such as Ne-OH. The mid-bond functions are placed at the half-distance between the Ne and the center of mass of OH and composed of {\em 3s3p2d2f2g}  functions with the following exponents: $sp$ 0.9, 0.3, 0.1 and $df$ 0.6, 0.2. The threshold for convergence of the total energies was set to 10$^{-10}$ E$_h$.

We calculated the $A'$ and $A''$ adiabatic PESs on a discrete grid of Jacobi coordinates $(R,\theta)$ with the OH  distance being frozen at $r_0=1.8509$ $a_0$, the average internuclear distance of the ground vibrational state.  The Jacobi radius $R$ is the distance of the Ne atom from the center of mass of  OH calculated with the masses of the most stable isotopes of O and H. The Jacobi angle $\theta$ describes the anisotropy of the PESs originating from the rotation of the Ne atom around the OH molecule (with $\theta=0^{\circ}$ corresponding to the collinear Ne--H-O arrangement). The discrete grid was composed of 52 radial points for  the $R$ variable and 19 angular $\theta$ points for the total of 988 points for each of the $A'$ and $A'$ PESs. The radial grid covered distances from 2.5 $a_0$ to 500 $a_0$ with variable step sizes $\Delta R=$ 0.1, 0.25, 1.0, 2.0, and larger). The $\theta$ grid was set up from $0^{\circ}$ to $180^{\circ}$ every 10$^{\circ}$. The largest radial distance of 500 $a_0$ was used to correct the RCCSD(T)-F12a interaction energies for the size consistency error, which arises due to the fact that the triple (T) corrections are not explicitly correlated. 

In order to perform scattering calculations we need to represent the discrete set of interaction PESs in the form of  analytical functions of the Jacobi coordinates $R$ and $\theta$. To this end, we fit the angular dependence of the diabatic half-sum and half-difference PESs $V_\text{sum}$ and $V_\text{diff}$ given by Eq.~(\ref{Vexpansion}) to a series of reduced Wigner functions $d_{\lambda 0}(\cos\theta)$ and $d_{\lambda 2}(\cos\theta)$, respectively. The resulting radial coefficients $V_{\lambda \mu}(R)$, $\mu=0,2$ were fitted using the reproducing kernel Hilbert space method of Rabitz and coworkers~\cite{ho:96} with the kernel describing dispersion-like terms for extrapolation. The resulting fit was smoothly joined with a long-range analytical expression based on the $C_n$ dispersion coefficients ($n=6{-}9$) obtained in a separate fit of the long-range part of the potential.

\begin{figure}[b]
\begin{center}
\includegraphics[width=1.45\linewidth, trim = 100 170 0 90]{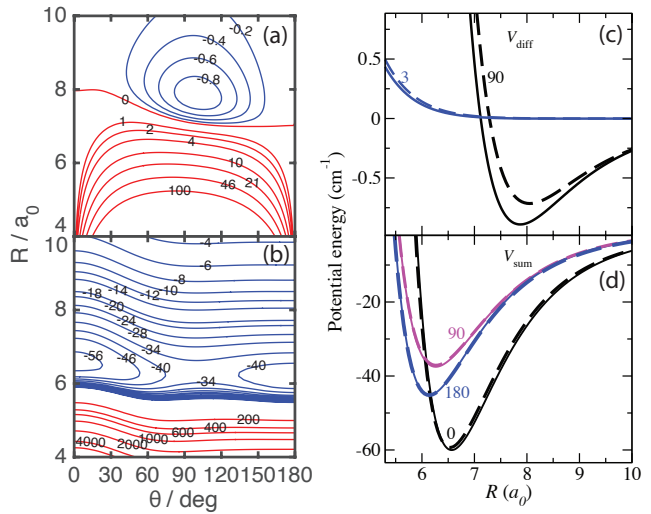}
\end{center}
\caption{Contour plots of the half-difference (a) and  half-sum (b) Ne-OH PESs calculated in this work as a function of the Jacobi coordinates $R$ and $\theta$ (in units of cm$^{-1}$).    (c)-(d) Radial dependence of the half-difference and half-sum PESs $V_\text{diff}(R,\theta)$ and  $V_\text{sum}(R,\theta)$ calculated in this work (full lines) and computed by Sumiyoshi {\it et al.}   \cite{Sumiyoshi:10} (dashed lines). The corresponding values of $\theta$ are indicated in degrees next to each curve.}
\label{fig:PES}
\end{figure}

Figures \ref{fig:PES}(a) and \ref{fig:PES}(b)  show the half-difference and half-sum  diabatic PES of the Ne-OH complex,  $V_\text{diff}(R,\theta) = \frac{1}{2}(V_{A''}-V_{A'})$ and $V_\text{sum}(R,\theta) = \frac{1}{2}(V_{A''}+V_{A'})$ as a function of $R$ and $\theta$. 
The global minimum of the  PESs is found in the collinear Ne--H-O geometry at $\theta_e=0^{\circ}$, $R_e=6.57$ $a_0$ and $D_e=60. 026$ cm$^{-1}$.  There is additionally a local minimum on the $A'$ PES, which occurs in a slightly skewed geometry at $\theta=66^{\circ}$, $R=5.85$ $a_0$, and $D_e=59.804$~cm$^{-1}$. The collinear Ne--O-H  minimum at $\theta=180^{\circ}$ is shallower with a well depth of 45.156 cm$^{-1}$ at $R=6.14$ $a_0$. This critical point is a minimum on the half-sum PES $V_\text{sum}$  and on the $A''$ adiabatic PES, but it is a saddle point on the $A'$ adiabatic PES. The T-shape region of the half-sum PES is characterized by a saddle point at $R=6.22$ $a_0$, $\theta=102^{\circ}$ and $D_e=-36.88$~cm$^{-1}$. This suggests that the Ne-OH complex has a fairly floppy structure with the Ne atom easily delocalized  between the two collinear minima. The half-difference  PES $V_\text{diff}$ is mostly repulsive with a shallow ($D_e\simeq1$~cm$^{-1}$) long-range minimum, as shown in Fig~\ref{fig:PES}(b).

Figures~\ref{fig:PES}(c) and (d) compare the radial dependence of our {\it ab initio} Ne-OH PESs with those computed by  Sumiyoshi {\it et al.} \cite{Sumiyoshi:10}. While the PESs are  overall very similar, small differences remain in the well depth and its location.
 In particular, the global minimum of our half-sum PES ($D_e = 60.026$~cm$^{-1}$)  is slightly deeper that that of the Ne-OH PES ($D_e = 59.4$~cm$^{-1}$) determined by Sumiyoshi {\em et al.}~\cite{Sumiyoshi:10}. This difference may originate from the slightly different values of the OH interatomic distance $r$ used in our calculations vs those of  Ref.~\cite{Sumiyoshi:10}. The latter used the equilibrium distance of OH ($r=r_e$), whereas we use the expectation value of the internuclear distance $r_0=\langle {v=0}|r|{v=0}\rangle$ in the ground rovibrational state of OH. Other numerical uncertainty factors resulting from the fitting procedures, and using the spin-unrestricted coupled-cluster approach by Sumiyoshi {\em et al.}, could also contribute to the discrepancies. The most sensitive to the level of {\it ab initio} calculations is the half-difference PES shown in comparison to the Sumiyoshi {\em et al.}'s PES  in Fig~\ref{fig:PES}(c).  The long-range shallow minimum in our half-difference PES is more attractive in the T-shaped Ne-OH geometry. These differences, albeit small, have a dramatic effect on cold Ne~+~OH scattering in the resonant regime as discussed in the main text.

To represent the  {half-sum and half-difference}  PESs in a form suitable for   quantum scattering calculations, we expand them in angular basis functions \cite{Alexander:82,Tscherbul:09}
\begin{align}\label{Vexpansion}\notag
V_\text{sum}(R,\theta) &= \frac{1}{2}(V_{A''}+V_{A'}) = \sum_{\lambda=0}^{\lambda_\text{max}}V_{\lambda 0}(R)P_\lambda(\cos\theta), \\
 V_\text{diff}(R,\theta) &= \frac{1}{2}(V_{A''}-V_{A'}) = \sum_{\lambda=2}^{\lambda_\text{max}}V_{\lambda 2}(R)d_{02}^\lambda(\cos\theta),
\end{align}
where $P_\lambda(\cos\theta)$ are the Legendre polynomials, and $d_{02}^\lambda(\cos\theta)$ are the reduced Wigner D-functions.

The radial coefficients $V_{\lambda 0}(R)$ and $V_{\lambda 2}(R)$ are evaluated using a 30-point Gauss-Legendre quadrature, and the expansions (\ref{Vexpansion})  are truncated at $\lambda_\text{max}=17$ to provide converged matrix elements of the interaction potential \cite{Tscherbul:09}.


\section*{Appendix B: Quantum scattering calculations}

The Hamiltonian of the Ne-OH collision complex is (in atomic units)  \cite{Alexander:82,Tscherbul:09}
\begin{equation}\label{HamiltonianNeOH}
\hat{H} = -\frac{1}{2\mu R}\frac{\partial^2}{\partial R^2}R + \frac{\hat{L}^2}{2\mu R^2} + \hat{V}(R,r,\theta) + \hat{H}_\text{mol},
\end{equation} 
where  $\mathbf{R}$ is  the vector pointing from Ne to the center of mass of OH, $\mathbf{r}$ is the internuclear separation vector in OH, and  $\theta$ is the angle between $\mathbf{R}$  and $\mathbf{r}$. Further, $\mu$ and $\hat{L}^2$ are the reduced mass and the orbital angular momentum for the collision, and 
\begin{equation}\label{Hmol}
\hat{H}_\text{mol} = \hat{H}_\text{rot} + \hat{H}_\text{SO} + \hat{H}_\Lambda + \hat{H}_\text{E} + \hat{H}_\text{B}
\end{equation}
 is the molecular Hamiltonian  consisting of the rotational, spin-orbit, $\Lambda$-doubling, Stark, and Zeeman terms \cite{Tscherbul:09}. To solve the Schr\"odinger equation  $\hat{H}|\Psi\rangle = E|\Psi\rangle$, we use the coupled-channel  (CC)  expansion of the wavefunction of the collision complex
\begin{equation}
|\Psi\rangle = \frac{1}{R} \sum_{\beta,\ell m_\ell} F_{\beta \ell m_\ell}(R)   |\beta\rangle |\ell m_\ell \rangle,
\end{equation}
 where $|\ell m_\ell\rangle$ are the eigenstates of $\hat{L}^2$ and $\hat{L}_Z$, and $\beta =|JM\Omega\Lambda\Sigma\rangle$ are Hund's case (a)  basis functions  for the internal degrees of freedom of OH \cite{Alexander:82}. These functions depend on the total angular momentum of the molecule $J$ and its projections on the space-fixed ($M$) and molecule-fixed ($\Omega$) quantization axes, and well as on the electronic basis functions $|\Lambda\Sigma\rangle$. The CC expansion leads to a system of CC equations  \cite{Tscherbul:09} for the radial solutions $F_{\beta \ell m_\ell}(R)$
\begin{multline}\label{CCsystem} 
\left[ \frac{d^2}{dR^2} + 2\mu E\right] F_{\beta  \ell m_\ell}(R) = 2\mu\sum_{\beta'\ell' m_\ell'}\langle \beta \ell m_\ell | \hat{V}(R,\theta,r) \\ + \frac{\hat{L}^2}{2\mu R^2} +\hat{H}_\text{mol} |\beta' \ell' m_\ell' \rangle F_{\beta'\ell' m_\ell'}(R).
\end{multline} 
which is integrated numerically using the log-derivative algorithm \cite{Manolopoulos:86} to produce the radial solutions $F_{\beta \ell m_\ell}(R)$. The application of the scattering boundary conditions to the $F_{\beta \ell m_\ell}(R)$ yields the $S$ and $T$-matrix elements as a function of collision energy and electric field \cite{Tscherbul:09,Pavlovic:09}. The  $T$-matrix elements are used to calculate the scattering amplitude [Eq. (2) of the main text], from which the ICS are computed by integration over the spherical polar angles of vector $\hat{R}$ ($\theta_R$ and $\phi_R$) using two-dimensional Gauss-Legendre quadratures. We verified that the resulting ICSs are the same as given by     Eq. (3)  of the main text.

The matrix elements of the Hamiltonian in Eq. (\ref{CCsystem}) are evaluated as described in our previous work \cite{Tscherbul:09}. The matrix elements of the molecular Hamiltonian $\hat{H}_\text{mol}$ are parametrized by the spectroscopic constants of  OH$(^2\Pi)$ \cite{Brown:03,Maeda:15}. The Ne-OH interaction is described by two adiabatic potential energy surfaces (PES) of $A'$ and $A''$ symmetries \cite{Alexander:82,Tscherbul:09} (see above).

The CC equations (\ref{CCsystem}) are integrated numerically using the modified log-derivative method \cite{Manolopoulos:86} on the radial grid extending from $R_\text{min}=2.5\, a_0$ to $R_\text{max}=75.02\, a_0$ with a grid step of $\Delta R = 0.02, a_0$ (for collision energies below 0.1~cm$^{-1}$). At higher collision energies, we use   $R_\text{max}=50.0\, a_0$ and a grid step of  $\Delta R = 0.05\, a_0$. The CC basis set included all total angular momentum states of OH with $J\le 4.5$ and partial waves with $\ell \le 7{-}9$ depending on the collision energy, and the following masses of  collision partners were used: $m_\text{Ne} = 19.9924401754$ and $m_\text{OH}=       17.002739$ amu. The spectroscopic constants of OH used to parametrize the molecular Hamiltonian (\ref{Hmol}) are the same as in our previous work on cold He~+~OH collisions \cite{Tscherbul:09,Pavlovic:09}. These references also contain further details regarding the evaluation of the matrix elements of the interaction potential starting from Eq.~(\ref{Vexpansion}).


 \begin{figure}[t]
\begin{center}
\includegraphics[width=0.65\linewidth, trim = 10 190 0 120]{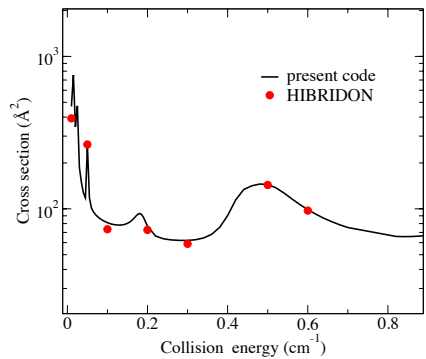}
\end{center}
\caption{ICSs for the $f\to e$ transition in Ne~+~OH($f,M=3/2)$ collisions  summed over all $M'$ states in the $e$ manifold vs. collision energy calculated using the HIBRIDON suite of scattering codes by Alexander {\em et al.}~\cite{Hibridon} and the code  developed in Ref. \cite{Tscherbul:09} (full line) at zero electric field. }
\label{figSM:xtest}
\end{figure}

At $R=R_\text{max}$ the log-derivative matrix is transformed to a basis, which diagonalizes the asymptotic Hamiltonian \cite{Tscherbul:09} and then matched to the Riccati-Bessel functions and their derivatives \cite{Johnson:73,Tscherbul:09} to yield the $S$-matrix elements, from which the scattering cross sections are obtained using standard expressions \cite{Johnson:73}. 
The calculated scattering cross sections are converged to $<10\%$ with respect to the basis set and radial grid parameters. 


To benchmark our quantum scattering calculations, we compared the ICSs calculated using an in-house $^1$S atom~+~OH quantum scattering code  developed in Ref.~\cite{Tscherbul:09}  with those computed using the publicly available general-purpose scattering   code HIBRIDON \cite{Hibridon}. Figure~\ref{figSM:xtest} shows the 
 ICSs for Ne~+~OH$(f,J=3/2,M=3/2)$ collisions as a function of collision energy at zero electric field. We observe excellent overall agreement, including near the low-energy scattering resonances, which serves as a verification  of our quantum scattering calculations.

\section*{Appendix C: Steric asymmetry for elastic scattering}
\begin{figure}[t!]
\begin{center}
\includegraphics[width=1.07\linewidth, trim = 30 360 0 220]{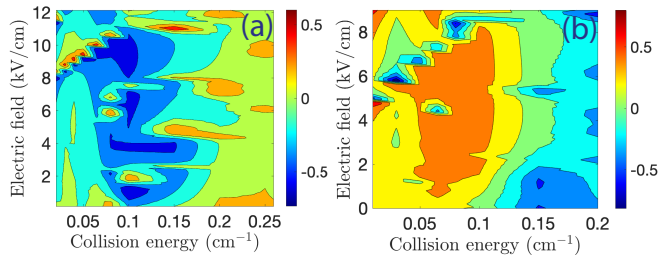}
\end{center}
\caption{Steric asymmetry calculated from the elastic ICSs  using the present Ne-OH PESs (left panel) and the PES of Ref.~\cite{Sumiyoshi:10} (right panel).}
\label{figSM:steric_asym_el}
\end{figure}

\vspace{-0.2cm}
Figure \ref{figSM:steric_asym_el} shows that elastic steric asymmetry defined as
\begin{equation}\label{steric_asym_el}
\mathcal{S}^\text{el} =  \frac{\sigma_{e,M}^\text{el} - \sigma_{f,M}^\text{el}}{\sigma_{e,M}^\text{el} + \sigma_{f,M}^\text{el} }
\end{equation}
for $M=3/2$.  Similarly to the inelastic steric asymmetry, this quantity measures the difference between the magnitude of the elastic ICSs for OH molecules colliding along vs. against the field direction. 

We observe that unlike its inelastic counterpart [see Eq. (4) of the main text] the elastic steric asymmetry is highly sensitive to the  details of the PESs used in scattering calculations over the whole range of collision energies and electric fields. This occurs due to the absence of the threshold suppression effects in elastic scattering.  As a result,  no universal regime exists for the stereodynamics of elastic atom-molecule collisions at low temperatures.  

%

\end{document}